\documentstyle[aps,epsf,prl]{revtex}
\begin{document}

\twocolumn[\hsize\textwidth\columnwidth\hsize\csname
@twocolumnfalse\endcsname
\draft
\title{Lithium-6: A Probe of the Early Universe}
\author{K. Jedamzik\cite{byline1}}
\address{Max-Planck-Institut f\"ur Astrophysik,
Karl-Schwarzschild-Str. 1, 85740 Garching, Germany} 
\maketitle

\begin{abstract}
I consider the synthesis of $^6$Li due 
to the decay of relic particles, such as 
gravitinos or moduli, after the epoch of 
Big Bang Nucleosynthesis. The synthesized 
$^6$Li/H ratio may be compared to $^6$Li/H 
in metal-poor stars which, in the absence 
of stellar depletion of $^6$Li, yields 
significantly stronger constraints on 
relic particle densities than the usual 
consideration of overproduction of $^3$He. 
Production of $^6$Li 
during such an era of non-thermal nucleosynthesis 
may also be regarded as a possible explanation for the 
relatively high $^6$Li/H ratios observed 
in metal-poor halo stars. 
\end{abstract}

%\pacs{PACS numbers: 04.70.Bw, 04.25.Dm, 97.60.Lf, 98.80.Cq}
\vskip2.2pc]

It is believed that most of the observed properties of the present
universe originate from out-of-equilibrium conditions during brief
periods in the evolution of the
very early universe. These may result during cosmic phase transitions,
an inflationary period followed by an era of reheating, 
and epochs of incomplete
particle annihilation. During such eras the production of 
\lq\lq unwanted\rq\rq\ relics
is also possible, and any
observational constraint on relics is important in order to limit a plethora
of proposed scenarios. 
Lindley~\cite{Li80} pointed out, that $\gamma$-rays present in the
primordial plasma below redshift
$z\, {}^<_{\sim}\, 10^7$ may spoil the agreement between light element
abundances synthesized during Big Bang Nucleosynthesis (hereafter, BBN)
and observationally
inferred primordial abundance constraints due to the possible
photodisintegration
of $^2$H. Similarly, 
possible overproduction
of mass two and three 
elements concomitant with the photodisintegration of $^4$He may result
from $\gamma$-rays injected below redshift $z {}^<_{\sim}\, 2\times 10^6$.
Such arguments have since been used to constrain the abundances 
of a variety of relics, such as massive decaying
particles~\cite{Mass}, 
and radiating cosmic
strings~\cite{Strings,SJBS95}, among others. 
In this letter I point out, that decay of 
relics after the
BBN era leads also to efficient production of $^6$Li.
%~\cite{remark1}. 
  
Injection of energetic $e^{\pm}$ and $\gamma$-rays into the primordial
plasma at high redshift due to the decay of a relic,
induce an electro-magnetic cascade on
the cosmic microwave background radiation (hereafter, CMBR) via pair production
of $\gamma$-rays on CMBR photons,  
$\gamma +\gamma_{\rm bb}\to e^- + e^+$, and inverse Compton
scattering of the produced pairs on the CMBR,
$e^{\pm} +\gamma_{\rm bb}\to e^{\pm} +\gamma$. This cascade is
halted only when $\gamma$-ray energies fall below the threshold for
pair production, i.e. for
$E_{\gamma} < E_C \simeq m_e^2/2E_{\rm bb}$. The resulting spectrum of 
\lq\lq breakout\rq\rq\ (i.e. $E_{\gamma} < E_C$) 
photons is quite generic, independent of the details of the injection
mechanism, and has been analyzed analytically and
numerically~\cite{Li80,A85,B90,P95,K95}. 
The number of \lq\lq breakout\rq\rq\ photons per unit energy interval is
well approximated by the following form~\cite{B90}
\begin{equation}
{n_{\gamma}(E_{\gamma})}
\approx \left\{ \begin{array}{ll}
K_0(E_{\gamma}/E_X)^{-1.5}
&  \mbox{for $E_{\gamma}<E_X$} \\ 
%[0.2in]
K_0(E_{\gamma}/E_X)^{-2}
&  \mbox{for $E_X<E_{\gamma}<E_C$} \\ 
%[0.2in]
\end{array}
\right.	
        \label{spectrum}
\end{equation}
where
$E_C(z)\approx 4.7\times 10^7z^{-1}$MeV, with $z$ redshift, and where
$E_X(z)\approx 1.78\times 10^6z^{-1}$MeV~\cite{P95}
represents a break in the spectrum. Here 
$K_0\approx {E_0/ (E_X^2[2 + {\rm ln}(E_C/E_X)])}$
is a normalization constant with $E_0$ the total energy in form of
electro-magnetically
interacting particles (i.e. $e^{\pm}$'s and $\gamma$'s with energies
well above $E_C$) injected by the decay. Subsequent interactions of the
\lq\lq breakout\rq\rq\ $\gamma$-rays are dominated by processes
on matter. The dominant process for energetic photons is
Bethe-Heitler pair production on protons and helium,
i.e.\, $\gamma + p\, ({}^4{\rm He})\to p\, ({}^4{\rm He}) + e^- + e^+$.
The cross section for this process is given by
\begin{equation}
\label{BetheHeitler}
\sigma_{BH}(E_{\gamma},Z)\approx {\alpha\over\pi}\sigma_{Th}\biggl(
{28\over 9}{\rm ln}\biggl[{2E_{\gamma}\over m_e}\biggr] - {218\over
27}
\biggr)Z^2\, ,
\end{equation}
for $1\ll E_{\gamma}/m_e\ll \alpha^{-1}Z^{-1/3}$, where $\alpha$
is the fine structure constant, $\sigma_{Th}$ the Thomson cross
section, $m_e$ the electron mass, and $Z=1$ for protons and 2 for helium. 
The pairs created
suffer inverse Compton scattering on CMBR photons
and generate a secondary generation of $\gamma$-rays which,
nevertheless,
is much softer than the \lq\lq breakout\rq\rq\ photons whose initial
spectrum is given by Eq.~(\ref{spectrum}).

Energetic $\gamma$-rays may also photodisintegrate $^4$He, i.e.
$\gamma +{}^4{\rm He}\to {}^3{\rm H}\, ({}^3{\rm He}) + p\, (n)$,
provided their energies are above the threshold for this process,
$E_{\gamma {}^4{\rm He}}^{th} = 19.81\,$MeV for production of $^3$H (which
decays into $^3$He)
and $20.58\,$MeV for direct production of $^3$He. Here the cross sections
for the production of $^3$H and $^3$He are almost equal. 
Deuterium may also result
from the photodisintegration process but with production typically
suppressed by a factor of ten compared to that of $^3$He~\cite{P95,SJBS95}.
A small
fraction of \lq\lq breakout\rq\rq\ photons photodisintegrate rather than
Bethe-Heitler pair produce. Thus, at redshift $z\, {}^<_{\sim}\,
2\times 10^6$ when $E_C\, {}^>_{\sim}\, 20\,$MeV cosmologically
significant production of $^3$He may result. The total number of $^3$H
nuclei produced by \lq\lq breakout\rq\rq\ photons is given by the
contribution from photons with different energies, $E_{\gamma}$~\cite{P95},
\begin{equation}
\label{tritium1}
N_{{}^3{\rm H}} =  \int_{E_{\gamma {}^4{\rm He}}^{th}}^{E_C}
{\rm d}E_{\gamma}{{\rm d}N_{{}^3{\rm H}}\over {\rm
d}E_{\gamma}}\, ,
\end{equation}
where
\begin{equation}
\label{tritium2}
{{\rm d}N_{{}^3{\rm H}}\over {\rm
d}E_{\gamma}} \approx {n_{\rm {}^4He}\,\sigma_{{}^4{\rm He}(\gamma ,p){}^3{\rm
H}}(E_{\gamma})\,n_{\gamma}(E_{\gamma})
\over
n_p\,\sigma_{BH}(E_{\gamma},1) +n_{\rm {}^4He}\,\sigma_{BH}(E_{\gamma},2)}\, ,
\end{equation}
and with $n_{\rm {}^4He}$, $n_p$ helium and proton density, respectively.
Eq.~(\ref{tritium2}) 
is essentially a computation of the probability
$P_{\gamma {\rm He}}$ that
a photon will photodisintegrate, given that its typical life time
towards pair-production is $\tau_{BH}\approx (c\sigma_{BH} n_p)^{-1}$.
It applies for $P_{\gamma {\rm He}}\ll 1$ and when $\tau_{BH}$ is
smaller than the Hubble time at the epoch of the cascade,
which is the case for $z\, {{}^>_{\sim}}\, 10^3$

It is important to realize that the photodisintegration process
leaves an initially non-thermal distribution of daughter nuclei,
which may participate in non-thermal nuclear reactions.
In fact, if only a small fraction of the produced energetic $^3$H and
$^3$He may further react via
${}^3{\rm H}\,({}^3{\rm He}) +{}^4{\rm He}\to {}^6{\rm Li} + n\, (p)$,
overproduction of $^6$Li may result. This reaction has 
energy threshold of $E_{{}^6{\rm Li}}^{th} = 4.80$ MeV for synthesis of
$^6$Li by $^3$H nuclei and
$4.03$ MeV by $^3$He nuclei. 
%and thus is of no importance during the BBN
%freeze-out where thermal nuclear reactions take place. 
The cross
section for $^3{\rm He}(\alpha ,p){}^6{\rm Li}$ has been measured at  
$32 + 3\,$mb at $28\,$MeV energy (in the lab frame)~\cite{K77}, where the first
contribution is from production of $^6$Li in the ground state and the
second is into the second excited state of $^6$Li (which
decays into the ground state). The cross section for $^3$H is expected
to be almost the same due to symmetry considerations. Unfortunately,
there seems to be no further experimental data
for this reaction. Theoretical calculations of nuclear reactions within
seven-nucleon systems~\cite{F91} 
may reproduce the experimental data and suggest
that the cross section is almost energy independent between threshold
and $E_{^3{\rm H}}\approx 35$ MeV. In contrast, $^6$Li production
during BBN has to proceed mainly via a reaction absent of an energy threshold,
${}^2{\rm H} + {}^4{\rm He}\to {}^6{\rm Li}$, 
with cross section in the $10 - 100$ nb range, a factor $\sim 10^6$
below that for $^3{\rm H}(\alpha ,n){}^6{\rm Li}$. BBN yields of
$^6$Li are therefore generically low $ 10^{-14}\, {}^<_{\sim}\, (^6{\rm
Li}/^1{\rm H})\, {}^<_{\sim}\, {\rm a\, few\,}\times 10^{-13}$~\cite{N97,V98}, 
with considerable uncertainty
remaining due to ill-determined reaction rates.

The kinetic energy transferred to the daughter $^3$H and $^3$He nuclei
during the photodisintegration process is a simple function of photon energy,  
$E_{^3{\rm H}}(E_{\gamma}) = (E_{\gamma}-E_{\gamma {}^4{\rm
He}}^{th})/4$ .
This implies that a
$\gamma$-ray needs $E_{\gamma}\approx 40$MeV in order to produce a
mass three nucleus sufficiently energetic to synthesize $^6$Li. The
main energy loss of energetic charged nuclei in the plasma is due to
Coulomb scattering off electrons and plasma excitations. 
Energy loss due to these processes per unit path length traveled 
by the nuclei is given by~\cite{J75}
$({{\rm d}E / {\rm d} x})_C = (Z^2\alpha / v^2)\,\omega_p^2\,{\rm ln}[
{\Lambda m_e v^2/ \omega_p}]$,
where $\omega_p^2 = 4\pi n_e\alpha /m_e$ is the plasma
frequency, $n_e$ the electron density,
$v$ the velocity of the energetic nuclei, $Z$ the charge of the nuclei, and
$\Lambda$ a factor of order unity. 
Given the above, it is 
possible to calculate the total $^6$Li yield resulting from
the energetic \lq\lq breakout\rq\rq\ photons. This is accomplished by
a convolution over (a) the initial $^3$H ($^3$He) energy which is given
by the spectrum of the photodisintegrating photons Eq.~(\ref{spectrum}), the
energy-dependent relative
reaction rates for $^4$He photodisintegration and Bethe-Heitler pair 
production Eq.~(\ref{tritium2}), and the relation between 
$E_{^3{\rm H}}$ and $E_{\gamma}$, and (b) the probability that a $^3$H
($^3$He) nuclei of given initial energy synthesizes with $^4$He to form
$^6$Li as it continuously looses energy by Coulomb interactions. 
This results in
\begin{eqnarray}
\label{convolve}
N_{{}^6{\rm Li}} = \int_{E_{\gamma {}^4{\rm He}}^{th} 
+ 4E_{{}^6{\rm Li}}^{th}}^{E_C}
{\rm d}E_{\gamma}{{\rm d}N_{{}^3{\rm H}}\over {\rm
d}E_{\gamma}} \quad\quad \nonumber \\
\times \int_{E_{{}^6{\rm Li}}^{th}}^{E_{^3{\rm H}}(E_{\gamma})}
{\rm d}E\, n_{{}^4{\rm He}}\,\sigma_{{}^3{\rm H}(\alpha
,n){}^6{\rm Li}}\,\biggl({{\rm d} x\over {\rm d}E}\biggl)_C\, .
\end{eqnarray}

Figure 1 shows the synthesized $^6$Li yield 
as a function of redshift per MeV of electro-magnetically
interacting energy injected by the decay of relics. The calculation
uses Eq.~(\ref{spectrum}) - (\ref{convolve}), 
as well as a $^4$He photodisintegration cross
section 
$\sigma_{{}^4{\rm He}(\gamma ,p){}^3{\rm
H}}\approx 0.8\,{\rm mb}\, (E_{\gamma}/40\,{\rm MeV})^{-2.9}$
applicable for $E_{\gamma}\, {{}^>_{\sim}}\, 40$ MeV~\cite{GH81}. 
The cross section
for $^3{\rm He}(\alpha ,p){}^6{\rm Li}$ is taken to be
energy-independent in the energy range of interest at a value of $38$ mb,
as suggested by the experimental determination and theoretical calculations. 
Given this, one finds 
that the contribution to the total yield of $^6$Li is peaked for
\lq\lq breakout\rq\rq\ photon energies of $E_{\gamma}\approx 70-80$ MeV,
resulting in $^3$H and $^3$He nuclei with energy $E_3\approx
15-20$ MeV, close to the energy where the cross section 
for $^3{\rm He}(\alpha ,p)^6{\rm Li}$ has been measured.
The calculation presented takes also account of two additional effects.
Tritium nuclei are unstable with half life $\tau_{{}^3{\rm
H}}=12.33$ y, such that
for redshifts below $z\, {}^<_{\sim} 10^4$ tritium nuclei may decay 
while still energetic enough to synthesize $^6$Li~\cite{remark2}.
An additional source of
$^6$Li is due to the photodisintegration of $^7$Li.
This has been included~\cite{remark3} and results in $^6$Li production
for redshifts $z\, {}^>_{\sim} 10^6$, as evident from Figure 1. 
At redshifts $10^6\, {}^>_{\sim}z\, {}^>_{\sim} 5\times 10^5$ the
$^6$Li yield may be somewhat uncertain due to the neglect of 
$\gamma\gamma$ scattering in the spectrum 
Eq.~(\ref{spectrum})~\cite{P95}. 

Dimopoulos {\it et al.}~\cite{D88} have shown that $^6$Li is also
synthesized in hadronic showers generated by the {\it hadronic} decay of
relics. Yields are dependent on the number and energy of injected
baryons in the decay,  which are sensitive to the mass and the
initial
decay products of the relic. They estimate a production of $\sim
5\times 10^{-11}$ of $^6$Li per MeV of relic particle energy
for hadronic decay of a particle with mass
$M_X\approx 1$ TeV. This is comparable to the yield from electro-magnetic
cascades for redshifts below $10^6$, but may dominate $^6$Li production at 
higher redshifts. Note that the $^6$Li yield presented here is
virtually independent
of decay mode, as it only depends on the fraction of relic energy
converted into electro-magnetically interacting energy. This fraction
is typically of order unity except for the case of \lq\lq
invisible\rq\rq\ decays.   

The abundance of $^6$Li has been determined for the
presolar nebula at $^6{\rm Li/H} \approx 1.5\times 10^{-10}$, and within the
atmospheres of hot, low-metallicity, Population II, halo stars. 
A number of authors have claimed detections of the $^6$Li/$^7$Li
ratio in the star HD 84937~\cite{HD,S98}, 
with the most recent observation yielding
$^6{\rm Li}/{}^7{\rm Li}= 0.052\pm 0.019$. Furthermore, a $^6$Li
detection
has also been claimed for the star BD $26\, 3578$~\cite{S98} at 
$^6{\rm Li}/{}^7{\rm Li}= 0.05\pm 0.03$. Both stars have 
metallicity of approximately $[{\rm Fe/H}]\approx -2.3$, and belong
to the Spite plateau of constant $^7$Li/H ratios, believed to reflect
the primordial $^7$Li abundance.
Recently, the first observational determination of $^6$Li/$^7$Li
ratios in the far more metal-rich ($[{\rm Fe/H}]\approx -0.6$) galactic
disk stars HD 68284 and HD 130551 has been claimed~\cite{N99}. 
Coincidentally, 
both stars have $^6{\rm Li}/{}^7{\rm Li}\approx 0.05$ with $^7$Li/H
ratios only slightly elevated from the Spite plateau.   
Given the value of the Spite plateau
$^7{\rm Li/H}\approx 1-2\times 10^{-10}$, one finds $^6{\rm Li/H}\approx
5-10\times 10^{-12}$, for the four stars where $^6$Li detections have
been claimed.

\begin{figure}[ht]
\epsfxsize=8.5cm
\hbox to\hsize{\hss\epsfbox{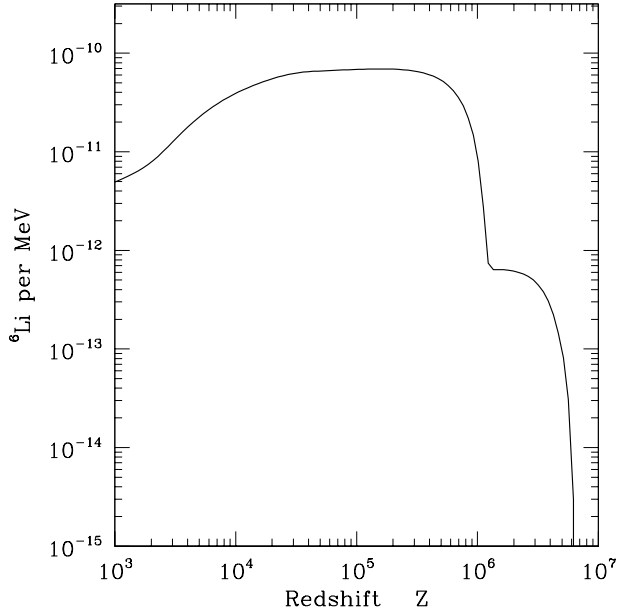}\hss}
\medskip
\caption[...]{\label{F1} Total number of $^6$Li nuclei produced
per MeV of energy in energetic electro-magnetically interacting
particles injected at an epoch with redshift $z$.}
\end{figure}

Whereas the origin of $^7$Li in hot, low-metallicity halo stars is known
to be primordial, $^6$Li,
as well as the isotopes $^{9}$Be, $^{10}$B (and some fraction of
$^{11}$B), are believed
to originate from spallation (${\rm p}, \alpha + {\rm CNO} \to {\rm
LiBeB}$) and fusion
($\alpha + \alpha \to {\rm Li}$) 
reactions of cosmic rays on interstellar gas. 
The abundances of these elements are expected to generically 
increase with increasing metallicity, since metallicity
represents a measure of the total \lq\lq action\rq\rq\ of galactic supernova
shock generated cosmic rays, up to the time of the formation of the
star. There is controversy as to the detailed
composition of the cosmic rays responsible for
LiBeB production. In order to explain an observed linear
relationship of Be versus Fe (which is contrary to what is
expected from \lq\lq standard\rq\rq\ cosmic rays with roughly
interstellar composition at the time of the supernova),
observationally allowed, but so far unknown, populations of
metal-enriched cosmic rays
have been postulated~\cite{D92}. Alternatively, it has been argued that if 
variation of O/Fe ratios with metallicity 
are taken into account, the observational
data of LiBeB may be reproduced~\cite{FO98}. 

\begin{figure}[ht]
\epsfxsize=8.5cm
\hbox to\hsize{\hss\epsfbox{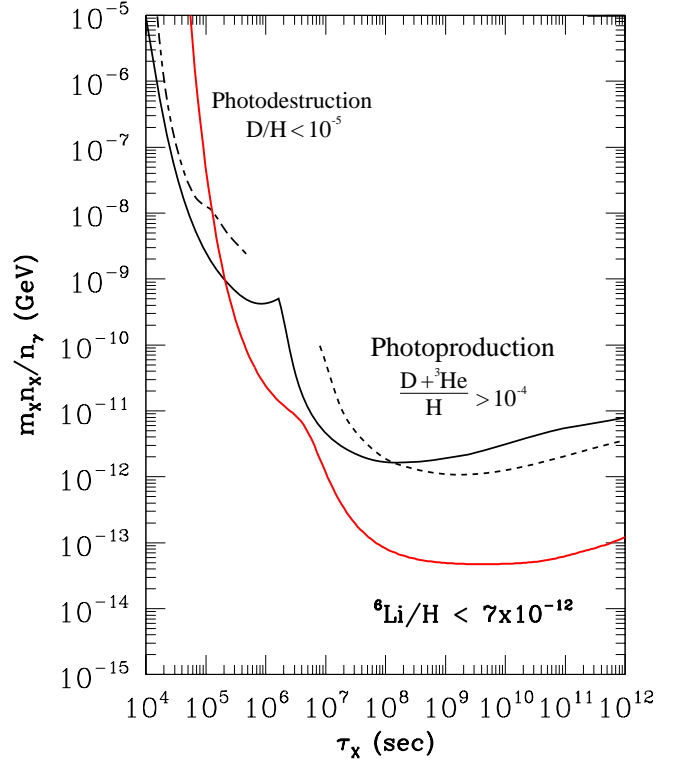}\hss}
\medskip
\caption[...]{\label{F2} Limits on the abundance of a relic, decaying 
particle with mass $m_X$, and density $n_X$, relative to photon density
$n_{\gamma}$, as a function of its lifetime $\tau_X$. 
The limits based on $^3$He and $^2$H 
destruction/production are taken from~\cite{Mass}.}
\end{figure}

The $^6$Li isotope may be depleted during the
pre-main sequence, as well as main sequence phase of stars. 
Nevertheless, the models by Ref.~\cite{V98} and~\cite{FO98}
have been used to argue against significant
(more than factor $\sim 2$) depletion of $^6$Li (and $^7$Li) in the
Pop II halo stars. The claim is, that by constructing models 
which reproduce the
solar system $^6$Li, the $^9$Be versus iron relation, as well as the
metallicity varying $^6$Li/$^9$Be ratios ($\simeq 80$ in Pop II stars, and
$\approx 5.9$ in the solar system), it seems not possible to
produce $^6$Li by far more than that observed in the halo stars
precluding significant astration of this isotope.
Ramaty {\it et al.}~\cite{R99} even claim, that whereas models
of metal-enriched cosmic rays may reproduce the $^9$Be data, {\it none} of
the existing cosmic ray models are able to synthesize $^6$Li in
abundance as observed in the halo stars, an argument which is based on 
cosmic ray energetics. 

In light of this it is intriguing to note that existing $^6$Li
observations, taken face value, are consistent with a 
\lq\lq no evolution\rq\rq\ hypothesis for metallicities below 
$[{\rm Fe/H}]\, {}^<_{\sim}\, -0.6$. Of course, it is most likely
that $^6$Li astration has occurred in the two disk stars with
metallicity $[{\rm Fe/H}]\,\sim\, -0.6$ ,
since cosmic ray nucleosynthesis scenarios predict a $^6$Li abundance
in excess of that observed in these stars~\cite{V98,FO98,R99}. 
However, in the absence of astration, 
a metallicity-independent abundance could be reconciled with a
primordial origin of $^6$Li, similar to the existence of a Spite
plateau for $^7$Li, though with $^6$Li originating from a very 
different process. A primordial origin could also offer an 
explanation
for the relatively high $^6$Li abundance in the Pop II stars.

One may use the observed $^6$Li abundance to derive a tentative limit
on the abundance of relics decaying after BBN, subject to the loophole
of $^6$Li astration in halo stars. In Figure 2, the demand of
pre-galactic $^6$Li synthesis not to exceed 
$^6$Li/H $\approx 7\times 10^{-12}$ was imposed on a relic decaying
with halflife $\tau_X\, {\rm ln}\, 2$.

In summary, I have shown that an era of non-thermal light-element
nucleosynthesis following the BBN freeze-out and
initiated by the electro-magnetic decay
of massive particles, evaporation of primordial black holes, or
radiating topological defects, not only leads to 
production of $^3$He and $^2$H as commonly known, but also results in
efficient $^6$Li synthesis. Here $^6$Li is mainly synthesized via $^3{\rm
H}(\alpha ,n)^6{\rm Li}$ by energetic tritium nuclei resulting from
the photodisintegration of $^4$He. This result provides additional
motivation for observations of $^6$Li in low-metallicity stars, accompanied by
an improved understanding of $^6$Li cosmic ray production, and stellar
depletion, since upper limits on the pregalactic abundance of
$^6$Li may be used to constrain non-equilibrium processes in the 
early universe. In the absence of stellar $^6$Li astration, current
observationally determined $^6$Li/H ratios in low-metallicity stars,
already provide a factor $\sim 50$ 
stronger constraint on the electro-magnetic decay
of relics than consideration of production of $(^2{\rm H} + {}^3{\rm
He})$ alone. 
This underlines the importance of the study of $^6$Li in metal-poor stars.
On the other hand, if a
relatively high plateau of $^6$Li/H ratios in low-metallicity stars
should be ever established, invaluable new insight in the evolution of the
very early universe might be gained. 

I wish to acknowledge many useful discussions with Jan B. Rehm.


\begin{thebibliography}{99}
\bibitem{remark2} The Coulomb
stopping time for a $^3$H with energy $E_{^3{\rm H}}$ is 
$\tau_{stop}^C\approx 2.0\times 10^7\, {\rm s}\, (z/10^4)^{-3}
(E_{^3{\rm H}}/15{\rm MeV})^{3/2}$. The decay product $^3$He may
still synthesize on $^4$He but looses energy more rapidly due to
its higher charge.
\bibitem{remark3} For direct production by $^7$Li photodisintegration
equations very similar to Eq.~(\ref{tritium1}) and
Eq.~(\ref{tritium2}),
with the appropriate replacements, have been used. Further, a cross
section for $^7$Li($\gamma , n$)$^6$Li as given in Ref.~\cite{DB88}, and
a BBN $^7$Li/H yield of $2\times 10^{-10}$ has been used. A similar
cross section for $^7$Be($\gamma, p$)$^6$Li has been assumed and 
$^7$Li($\gamma, p$)$^6$He has been, conservatively, neglected.
\end{thebibliography}
\end{document}